# Accomplishments in genome-scale *in silico* modeling for industrial and medical biotechnology

Caroline B. Milne[1,2], Pan-Jun Kim[1], James A. Eddy[1,3], and Nathan D. Price[1,2,4]

[1] Institute for Genomic Biology, [2] Department of Chemical and Biomolecular Engineering, [3] Department of Bioengineering, [4] Center for Biophysics and Computational Biology, University of Illinois, Urbana, IL 61801, USA

**Correspondence to:**

Nathan D. Price
Department of Chemical and Biomolecular Engineering
MC-712, Box C-3
600 South Matthews Avenue
Urbana, IL 61801

Tel.: +1 217 244 0596
Fax: +1 217 333 5052
Email: ndprice@illinois.edu

**Abbreviations**

GEM, genome-scale model
FBA, flux balance analysis
rFBA, regulatory FBA
dFBA, dynamic FBA
MOMA, minimization of metabolic adjustment
TRN, transcriptional regulatory network
LAB, lactic acid bacteria

## ABSTRACT

Driven by advancements in high-throughput biological technologies and the growing number of sequenced genomes, the construction of *in silico* models at the genome scale has provided powerful tools to investigate a vast array of biological systems and applications. Here, we review comprehensively the uses of such models in industrial and medical biotechnology, including biofuel generation, food production, and drug development. While the use of *in silico* models is still in its early stages for delivering to industry, significant initial successes have been achieved. For the cases presented here, genome-scale models predict engineering strategies to enhance properties of interest in an organism or to inhibit harmful mechanisms of pathogens or in disease. Going forward, genome-scale *in silico* models promise to extend their application and analysis scope to become a transformative tool in biotechnology. As such, genome-scale models can provide a basis for rational genome-scale engineering and synthetic biology.

## 1 INTRODUCTION & BACKGROUND

Genome-scale *in silico* models provide a powerful resource to guide rational engineering of biological systems for applications in industrial and medical biotechnology. An accurate **genome-scale model (GEM)** can help predict the system-wide effect of genetic and environmental perturbations on an organism, and hence drive metabolic engineering experiments. Since the development of the first GEM in 1999 (*Haemophilus influenza* [1]), systems modeling approaches have worked towards efficiently utilizing increasingly available high-throughput biological data (e.g., genomics, transcriptomics, proteomics, metabolomics) to bring genomes to life. An important challenge in this field is to enable the rapid development of predictive computational models for any sequenced organism by harnessing these high-throughput experimental technologies. The compelling need for this ability is evidenced by the gap between the number of sequenced organisms and corresponding GEMs (**Figure 1**).

Three classes of networks have been reconstructed in biochemical detail and converted into GEMs. **Metabolic GEMs** quantify a cell's metabolic potential, and thus allow researchers to probe new phenotypes and network states [2]. **Transcriptional regulatory networks (TRNs)** describe transcription-factor-promoter interactions and associated environmental influences to provide information about environment-specific enzyme activity. As such, TRNs can be fused with metabolic GEMs to form more predictive integrated metabolic-regulatory network models [3-6]. The newest genome-scale network type, **transcriptional-translational models** [7]**,** captures information passage from DNA to RNA to proteins. Each network-based GEM is built upon a stoichiometric formalism allowing for the mathematical representation of biochemical information (see [2, 8-10]). The present review focuses heavily on metabolic GEMs because they are the most commonly formulated and span a broad range of applications.

Numerous constraint-based methods are available to explore the phenotypic potential of the three GEM types, and by extension the associated biological system. To assist in understanding the case studies discussed herein, we briefly summarize some of these procedures (see [11] for review). In constraint-based analysis, physico-chemical and environmental constraints are applied as balances (e.g., mass, energy) and bounds (e.g., flux capacities, thermodynamics). These constraints define a solution space describing all possible functions (allowable phenotypes) of the system. **Flux balance analysis (FBA)** determines the distribution of reaction fluxes that optimize a biological objective function (e.g., biomass, ATP) [12, 13]. This capability is particularly useful for simulating the effect of genetic perturbations (e.g., gene knockouts or mutations) on the resulting metabolic phenotype. Two extensions of traditional FBA, **regulatory FBA (rFBA)** [3, 6] and **dynamic FBA (dFBA)** [14], enable analysis and hypothesis generation where external metabolite concentrations and environmental conditions vary with time. **Minimization of metabolic adjustment (MOMA)** assumes that, after a mutation, the organism seeks to minimize the total metabolic change relative to the wildtype (unlike FBA, which assumes a new optimized network state) [15]. **OptKnock** [16] is a computational procedure used to design strains with enhanced capabilities by identifying gene deletions that align the cellular objective (e.g., growth) with the engineering objective (e.g., biofuel production). The effects of gene additions from related organisms can be included in an analogous fashion using **OptStrain** [17].

This review provides detailed examples of how constraint-based GEM analysis has been used for a broad range of applications in industrial and medical biotechnology (**Figure 2**). To date, there are over 50 organism-specific GEMs (**Table 1**) that have been surprisingly successful in predicting cellular behavior (e.g., the effects of gene deletions on growth or secretion rates). In biotechnology applications, GEMs are commonly used to guide enhancement of a particular property of interest (e.g., biofuel or pharmaceutical production) or to better understand systemic behavior. Hence, two specific uses for GEMs are addressed: (i) elucidation of the global properties of network structures and (ii) constraint-based modeling for predicting the phenotypic effects of genetic and environmental perturbations.

## 2 INDUSTRIAL BIOTECHNOLOGY APPLICATIONS OF GENOME-SCALE *IN SILICO* METABOLIC MODELS

Metabolic GEMs provide a valuable tool to harness microorganisms as cell factories to sustainably produce chemicals and pharmaceuticals. The ability to integrate targeted modifications within the context of the whole organism helps model-guided approaches to minimize undesired secondary effects. An iterative model generation, hypothesis formation, and model refinement process is central to the systems biology approach (**Figure 3**). Current metabolic GEMs for industrial biotechnology fall into four main application categories: food production, biopolymers, biofuels, and bioremediation.

### 2.1 Food Production and Engineering

In the food and beverage industry, metabolic GEMs have been created to explore and improve fermentation byproduct formation by lactic acid bacteria (LAB). In addition to lactate, LAB produce bacteriocins, exopolysaccharides, polyols, B vitamins and compounds that affect food texture, taste, and preservation [18].

***Lactobacillus plantarum*** is used in industrial food fermentations and advertized as a probiotic organism. FBA was used to compare the typical estimation method for ATP production (based on lactate and acetate formation) to that predicted by the metabolic GEM, and was found to match. The accuracy of the acid-formation based method had been questioned because some inputs to lactate and acetate formation do not yield ATP. During the ATP production analysis, it was discovered that transamination of aromatic and branched chain amino acids contributes to ATP production. A second investigation with the GEM investigated the discrepancy between experimental and FBA predicted growth rates and lactate formation. FBA predicted mixed acid production (primarily acetate, ethanol, formate) when optimized for growth, while homolactic fermentation is observed experimentally. Additionally, the FBA-predicted growth rate was higher than expected. These differences were thought to result from the experimental observation that *L. plantarum* uses an ATP inefficient route for lactate production, and thus does not maximize ATP production as its cellular objective (the FBA assumption used) – likely stemming from its evolution in nutrient-rich environments. This observation was investigated further in a study that evolved an experimental strain for growth on glycerol [19]. The poor substrate expectedly forced the strain into optimization for growth, producing mainly lactate with an experimental growth rate of $0.26h^{-1}$, compared to $0.324h^{-1}$ found *in silico*. Thus, the experimental mutant developed to follow traditional FBA assumptions agreed better with *in silico* predictions.

In addition to typical LAB production applications in the food industry, ***Lactococcus lactis*** has applications relating to the *in situ* production of flavor, texture and health contributing food components. The GEM for *L. lactis* was used to predict modifications for enhanced production of diacetyl, a flavor compound in dairy products [20]. FBA and MOMA were used to optimize for production of the intermediate 2-acetolactate. *In silico* predictions starting with a known mutant strain yielded an additional deletion for increased acetate formation. In a subsequent deletion study on the new mutant, three more gene deletions predicted a redirect of carbon flux to 2-acetolactate production. Another application of *L. lactis* has been as an oral delivery vehicle for recombinant protein vaccines. To investigate this, the *L. lactis* GEM was updated to include recombinant protein synthesis reactions and used to optimize production of recombinant proteins [21]. Specifically, this study optimized production of Green Fluorescent Protein (GFP) (a model heterologous protein) using dFBA. The top performing strain predictions were tested *in vivo* and found to have 15% increased GFP production. The increase in expression was lower than predicted using the GEM, however the qualitative effect was still observed.

***Streptococcus thermophilus*** is commonly used in the production of yogurt and cheeses involving high cooking temperatures. The metabolic GEM enabled the comparison of *S. thermophilus* with *L. plantarum* and *L. lactis* metabolism [22]. Considering its evolution in protein-rich milk environments, *S. thermophilus* was surprisingly found to produce 18 amino acids. The GEM also identified a unique acetaldehyde (yogurt flavor) production pathway.

**2.2 Production of Biopolymers**

Today, most synthetic materials (e.g., plastics) are produced via petroleum refining.

In an effort to reduce dependence on unsustainable processes, alternative production routes for plastics are desirable. For example, poly-3-hydroxyalkanates (PHAs) are microbial produced biodegradable polyesters that could potentially replace petrochemical-based plastics.

PHA production was investigated using two metabolic GEMs of *Pseduomonas putida*. The first GEM was used to suggest precursor metabolites and showed that select fatty acids and carbohydrates were the best PHA precursors [23]. This was expected since carbon sources leading to high levels of acetyl coenzyme A (acetyl-CoA) are good PHA production candidates. Soon after the publication of this *P. putida* model, a second metabolic GEM [24], also analyzed to improve PHA production, was published. PHA and biomass (growth) pathways utilize the same metabolic precursors, so FBA predicts no PHA production when optimizing for growth. To overcome this, OptKnock was applied to the second GEM to increase the pool of the primary precursor acetyl-CoA. Six mutations were predicted, one of which demonstrated a 29% acetyl-CoA increase.

### 2.3 Production of Biofuels

Biofuels have potential to provide a sustainable and environmentally-friendly fuel source. Metabolic GEMs hold great promise to guide strain design for improved biofuel production by microorganisms [25]. In addition to the model fermentation organisms for ethanol, GEMs for lesser-characterized organisms that naturally exhibit useful properties are attractive for biofuel production. Currently, GEMs to improve ethanol, butanol, hydrogen and methane production have been developed and studied. The hydrogen producing algae *Chlamydomonas reinhardtii* is discussed in Section 2.5.

Acetone-butanol-ethanol production

For alcohol production, a global understanding of metabolic behavior is critical. Microbe production of alcohols is limited by the toxicity of these compounds at high concentration. Understanding the solution space defined by the metabolic network reveals whether the organism has reached its maximum production potential and is limited by toxicity (requiring engineering approaches that delve into dynamics and regulation). If not, the stoichiometric threshold has not been reached and alcohol production can be enhanced by redistributing carbon flux.

The *E. coli* and *S. cerevisiae* models have been used to improve ethanol production and are discussed in Section 4. Similarly, ***Clostridium acetobutylicum*** – the natural acetone-butanol-ethanol production organism that advantageously co-ferments pentoses and hexoses – has two GEMs [26, 27] that can be used to increase biobutanol production. In [26], 207 lethal reactions were found on minimal media, 140 on partially supplemented medium and 85 on supplemented medium [27] found 194 essential reactions.

<u>Methanogens</u>

Methanogens anaerobically convert low-carbon substrates to methane, and can degrade industrial, agricultural and toxic wastes containing large amounts of organic material. A GEM was reconstructed for ***Methanosarcina barkeri*** to study methanogenesis, representing the first archaeon GEM [28]. This model led to 55 new functional genome annotations, was used to suggest a minimal media, and uncovered the stoichiometry of three previously uncharacterized aspects of methane production.

Mutualistic methane production was investigated in the coupled study of ***Desulfovibrio vulgaris*** and ***Methanococcus maripaludis*** metabolic behavior, the first demonstration of a flux balance model for a two-organism system [29]. Though not genome-scale, it represents an interesting application of traditional constraint-based analysis. The two-system model was developed by separately reconstructing the central metabolism of *D. vulgaris* and *M. maripaludis*, and then integrating the networks as a single syntrophic system by compartmentalization. Unlike in eukaryotic models, compartments were separated by the extracellular environment, making transporter existence in both species critical. From this model, it was discovered that formate was not required as an electron shuttle between the organisms, but that growth was not possible without hydrogen transfer.

### 2.4 Applications in bioremediation

Bioremediation takes advantage of a microbe's ability to reduce and potentially eliminate toxic effects of environmental pollutants. Additionally, microbes capable of degrading harmful waste produce useful chemicals as byproducts, and hence are intriguing production organisms as well [30].

***Acinetobacter baylyi*** is an innocuous soil bacterium that degrades pollutants and produces lipases, proteases, bioemulsifiers, cyanophycine, and biopolymers. *A. baylyi* is easily transformed and manipulated by homology-directed recombination, enabling straightforward metabolic engineering. Therefore, the GEM is accompanied by an extensive library of mutants, and was validated against wildtype growth phenotypes in 190 environments and gene essentiality data for nine environments [31].

***Geobacter metallireducens*** reduces Fe(III) and is used in bioremediation of uranium, plutonium, technetium, and vadium. Its ability to produce electrically conductive pili makes it useful for harvesting electricity from waste organic matter and as a biocatalyst in microbial fuel cell applications. Using *G. metallireducens*' GEM, growth on different electron donors and electron acceptors was investigated [32]. Model analysis revealed energy inefficient reactions in central metabolism, and experimental data suggested that the inefficient reactions were inactive during biomass optimization on acetate, but up-regulated when grown with complex electron donors. Additionally, the model was tested for flux predictions by comparison with $^{13}C$ labeling flux analysis. Simulations suggested the TCA cycle was used to oxidize 91.6% of acetate, in agreement with 90.5% in $^{13}C$ labeling experiments.

***Geobacter sulfurreducens*** has similar industrial applications to *G. metallireducens*. OptKnock was applied to the *G. sulfurreducens* GEM [33] to improve extracellular electron transport [34]. Gene deletions in the fatty and amino acid pathways and in central metabolism were predicted to increase respiration and cellular ATP demand. To study the ATP demand increase, an ATP drain was added to the GEM. The model showed the rise in ATP usage correlated to decreased biomass flux and increased respiration rate. Experimental results confirmed that an ATP drain demonstrates the predicted results. Increasing electron transfer in *G. sulfurreducens* has advantages in both bioremediation and microbial fuel cell development, though increased fuel cell current was not found with this mutant strain.

## 2.5 Photosynthetic Organisms

The sun's energy can be captured either directly by using photosynthetic organisms as cell factories, or indirectly through plant biomass. Photosynthetic organisms can (i) remove $CO_2$ from the environment, thereby reducing the impact of global warming; (ii) use light to produce carbon-based products; and (iii) create energy gradients. While there is one plant GEM available (Nature Precedings [35]), this section will focus on photosynthetic microbes.

The algae ***Chlamydomonas reinhardtii*** is most commonly utilized for biofuel and biohydrogen production. *C. reinhardtii*'s GEM was reconstructed using an iterative method that integrates experimental transcript verification with computational modeling [36]. An initial metabolic network revealed genes needing experimental definition and validation, the completion of which refines the model through verification of hypothetical transcript existence. Resulting pathway gaps were filled by incorporating alternative enzymes, providing the basis for further transcript verification and network modeling.

***Halobacterium salinarum*** is an extreme halophilic archaeon capable of surviving with light as its only energy source. It produces bacteriorhodopsin (a light-driven proton pump) – the only known structure with non-chlorophyll based photosynthesis

– for use in optical security, optical data storage, and hologram creation. *H. salinarum* can also store energy (like a battery) using a large potassium gradient. Its GEM [37] was used to investigate aerobic essential amino acid degradation, and to integratively study energy generation, nutrient utilization, and biomass production.

Cyanobacteria are a subset of prokaryotes that execute oxygenic photosynthesis. ***Synechocystis sp.*** is a fresh water cyanobacterium for which powerful genetic tools are available (e.g., transformation tools, genetic markers). As a potential biofuel production organism, *Synechocystis* could convert $CO_2$ to carbon-based products. To test this ability, two genes were experimentally transformed into the metabolic network of *Synechocystis* to complete an ethanol-producing pathway. The $CO_2$ fixation to pyruvate was diverted to ethanol production – allowing for direct conversion of $CO_2$ to ethanol using only light energy. To investigate the added pathway's systemic effects, the two reactions corresponding to the gene additions were added to the GEM [38]. Analysis showed that the mutant strain should also now produce succinate and malate, as was subsequently verified experimentally.

## 3 MEDICAL BIOTECHNOLOGY APPLICATIONS OF GENOME-SCALE *IN SILICO* METABOLIC MODELS

In addition to applications in industrial biotechnology, systems-level metabolic modeling has been widely utilized in medical biotechnology. To capture the potential of constraint-based analysis and further improve drug production and target identification, metabolic GEMs spanning a range of diseases have been formulated. Demonstrated applications are grouped into three categories: anti-pathogen target discovery, drug and nutrient production, and mammalian systems.

### 3.1 Anti-pathogen Target Discovery

Microbial strains are the causative agents of numerous human diseases. Pathogen GEMs are thus primarily used to identify drug targets that would inhibit cellular function. Importantly, the GEM for humans [39, 40] informs these pathogen studies by identifying enzyme targets essential for the pathogen and not for humans. Most modeling studies of pathogens generate sets of essential genes and reactions under conditions representing their host environment to identify new antibiotic targets (**Figure 4**). A smaller number of studies report potential chemical inhibitors of these targets, and models have even been used to predict the specific effects of various drug compounds on the organism.

Modeling of ***Staphylococcus aureus***, a bacteria infecting multiple regions of the body, aims to elucidate the origin of its antibiotic resistance and to identify new drug targets. Its first metabolic GEM was used to identify essential genes and reactions on both rich and minimal media [41]. In this study, the authors generated a literature-derived list of potential combative drugs (chemical inhibitors corresponding to essential reaction targets) for the predicted targets. A later study identified *metabolites* essential for *S. aureus* survival [42]. A second GEM was extensively validated against experimental data and used to predict 158 lethal intracellular reaction knockouts [43]. Five of these knockouts had already been experimentally identified as lethal. Further analysis showed that biosynthesis pathways for glycans and lipids were particularly susceptible to deletions, making them interesting for antibiotic development. The most recent *S. aureus* modeling study combined metabolic reconstruction methods with genomic and sequence homology data to build a set of models representing the 13 different *S. aureus* strains [44]. 44 genes were predicted to be unconditionally essential across all strains. While a number of the essential genes were reported to have roles in fatty acid biosynthesis, the majority of the 10 common synthetic-lethal gene pairs identified belong to amino sugar biosynthesis pathways.

Respiratory Pathogens

***Haemophilus influenzae*** causes otitis media as well as acute and chronic respiratory infections, most often in children. Even with the *H. influenzae* type b (Hib) vaccine, an estimated 380,000 to 600,000 Hib deaths sill occur world-wide each year.

Furthermore, non-typeable *H. influenzae* strains lacking the vaccine target are becoming a major pathogen in both children and adults. The *H. influenzae* GEM was initially used to identify 11 genes predicted as critical in minimal substrate conditions [1]. Interestingly, six of the 11 genes were also determined to be critical in more complete substrate conditions reflecting the human host environment of *H. influenzae*. A later study integrated protein expression data with the model to predict essential enzymatic proteins in aerobic and microaerobic conditions [45].

***Mycobacterium tuberculosis*** is a significant cause of human disease in the third world, killing over two million people annually. Two metabolic GEMs for *M. tuberculosis* exist: GSMN-TB [46] and *i*NJ661 [47]. The GSMN-TB model contains five genes encoding enzymes that are known drug targets, all correctly predicted to be essential [46]. In a later study, FBA was combined with gene expression data to interrogate the metabolic network and predict the effects of different drugs, drug combinations, and nutrient conditions on mycolate biosynthesis [48]. Mycolates are key components of the mycobacterial cell wall, and mycolate metabolism is a target of well-known antituberculosis drugs. A separate study using *i*NJ661 identified mycolate as an essential metabolite [49]. Combined, these results suggest that mycolate biosynthesis and degradation pathways are viable targets for new drug discovery. Applying sampling and flux coupling methods to *i*NJ661, 50 known TB drug targets were mapped to hard-coupled reaction (HCR) sets, where a single drug target knocks out an entire set's functionality [47]. Terminating the activity of other enzymes in an HCR theoretically has the same effect, suggesting novel targets. Most recently, gene and reaction essentiality results obtained from both GEMs were integrated into a larger *in silico* target identification pipeline for *M. tuberculosis* that incorporates protein-protein interaction network analysis, experimentally derived essentiality data, sequence analyses, and structural assessment of targetability [50].

Another important respiratory pathogen studied through genome-scale modeling is ***Pseudomonas aeruginosa***. The ability of *P. aeruginosa* to form biofilms in low oxygen environments allows it to chronically infect the lungs of cystic fibrosis patients. *P. aeruginosa* is also responsible for nosocomial infections and acute infections in immunocompromised patients. *In silico* gene deletions performed with its metabolic GEM showed strong agreement with published knockout data [51].

Gastrointestinal Pathogens

***Helicobacter pylori*** targets the gastric mucosa, leading to diseases such as gastritis, peptic ulceration and gastric cancer. Seven essential genes were predicted in the initial *H. pylori* GEM under four test conditions, representing varying aerobic levels and nutrient availability [52]. Importantly, the overall variation between conditions revealed that gene essentiality is dependent on the *in silico* environment. Using an updated *H. pylori* GEM, a later study identified 128 essential genes, and the results were validated using published experimental data [53]. Most essential genes predicted belonged to either the cell wall or vitamin and cofactor subsystems. In a study predicting essential metabolites for cell growth [42], meso-2,6-diaminoheptanedioate was confirmed as a potential target, while ADP-d-glycero-d-manno-heptose was identified as a potentially novel target.

***Salmonella typhimurium*** is a source of human gastroenteritis and causes systemic infection in mice studied as a model for human typhoid fever. A variation of the typical target-prediction approach identified potential strategies for vaccine development [54]. Specifically, gene expression data was used to infer the host environmental conditions to which *S. typhimurium* might be exposed during infection. Model simulations predicted genes essential for intracellular survival, providing potential targets for generating avirulent attenuated strains for vaccines. A second *S. typhimurium* GEM showed good agreement between simulation and experimental results for growth patterns under different substrate conditions [55].

Pathogens Infecting Other Systems

***Neisseria meningitidis*** causes meningitis and meningococcal septicemia, and is classified into serogroups (groups containing a common antigen) A, B and C.

Serogroup B is common in developed countries and has no vaccine. The membrane protein PorA has been identified as a major inducer of, and target for, bactericidal antibodies. As genetically engineered strains expressing more than one PorA subtype are now being produced, GEMs can aid in process development of the cultivation step. The metabolic GEM was therefore used to define a minimal medium for *N. meningitidis* growth (successfully tested in batch and chemostat cultures) [56].

***Yersinia pestis*** infects the lymphatic system and causes bubonic plague, a disease without a vaccine that still affects thousands of people annually. The metabolic GEM was used to identify 74 lethal gene deletions and 39 synthetic lethals [57]. Similarly, *in silico* gene deletion studies on the ***Leishmania major*** GEM [58], the first GEM for a protozoan, revealed multiple essential genes (e.g., trypanothione reductase encoding genes) that are absent in humans. *L. major* is the causative agent of cutaneous leishmaniasis in mammalian hosts and is similar to other *Leishmania* species causing diffuse cutaneous, mucocutaneous and visceral forms of the disease.

***Mycoplasma genitalium*** is the closest known representation of the minimal gene set required for bacterial growth. Additionally, *M. genitalium* is sexually transmitted and causes nongonococcal urethritis in men, genital tract inflammatory diseases in women, and is thought to increase the risk of HIV-1 contraction. Presently, model predictions have helped to identify minimal media growth components [59].

Human oral pathogens such as ***Porphyromonas gingivalis*** are the leading cause of carious and periodontal disease. Lipopolysaccharides (LPS) present in the bacterial outer membrane trigger the human immune system. The *P. gingivalis* GEM identified several gene deletions preventing LPS production [60]. One predicted strain was confirmed to suffer negative effects, though it was still viable. Blocking LPS production would allow for control of the negative inflammatory responses.

## 3.2 Production of Drugs and Nutrients

Of interest, some microbial organisms produce antibiotics and other compounds conveying health benefits to humans (e.g., vitamins). Analysis of metabolic GEMs for both traditional and novel drug production microorganisms serves to improve production efficiency and assist in identifying new drug production routes.

### Nutrients & Dietary Supplements

***Corynebacterium glutamicum*** is used industrially to produce amino acids, particularly L-lysine and L-glutamate, and can produce ethanol and organic acids under oxygen deprivation conditions. The first metabolic GEM assisted in prediction of targets for improved lysine production, showing that lysine production via direct dehydrogenase gives a higher product yield [61]. Soon after publication of the first GEM, a second metabolic GEM was published and used to find candidate gene deletions to increase organic acid production under oxygen deprived conditions [62]. Improving lactate production required interruption of succinate-producing reactions. Disruption of oxidative phosphorylation reactions also predicted improved production of lactate because NADH oxidation demand increased. Finally, reactions in the pentose phosphate pathway were predicted to increase lactate production because an alternative reaction was needed to produce NADPH (malate to pyruvate) and the increase in pyruvate was converted to lactate. Succinate production was also predicted to be improved by interrupting the lactate producing reactions.

<u>Pharmaceuticals</u>

Aside from *E. coli,* ***Bacillus subtilis*** is one of the best-characterized prokaryotes. Its ability to produce antibiotics, high quality enzymes and proteins, nucleosides, and vitamins makes it an important industrial organism. Two genome-scale metabolic models have been created, the second of which used the SEED annotation [63, 64]. Analysis showed that 79% of the reactions from the earlier model were present in the later model, with 64% agreement in gene-reaction mapping. The newer model contains a larger number of reactions due to improved annotation and more detailed characterization of biomass composition.

***Streptomyces coelicolor*** also produces antibiotics, as well as secondary metabolites such as immunosuppressants and anti-cancer agents. It has been demonstrated experimentally that increasing the supply of primary metabolites (those directly involved in cell function and growth) – via decreased flux through primary metabolic pathways – leads to increased secondary metabolite production in various *Streptomyces* strains. The *S. coelicolor* GEM [65] was used to study the effect of reduced phosphofructokinase (PFK) activity on increasing secondary metabolite production [66]. The model subset used showed that applying constraints to limit secretion of other secondary metabolites (acetate, acetaldehyde, ethanol, formate, and proline) did not increase antibiotic production when PFK activity was reduced. The predicted decrease in specific growth rate and increase in pentose phosphate pathway flux was observed experimentally. Another study applied flux variability analysis to the original metabolic GEM to investigate the effects of different culture feed conditions on glucose assimilation and antibiotic production [67].

### 3.3 Mammalian Systems

Metabolism is a critical aspect of human physiology, and metabolic malfunction is a major contributing factor in many human diseases. Metabolic modeling of mammalian cells can be used to study tissue specific function [68] and human disease [69, 70]. Mammalian cell cultures (non-human) can also be used in the production of biopharmaceuticals such as monoclonal antibodies and vaccines [71].

The recent completion of a global reconstruction of the metabolic network in ***Homo sapiens*** [39, 40] represents a significant milestone in human systems biology. In addition to the typical network capabilities determined by constraint-based modeling, Human Recon 1 has enabled analysis of relationships between network topology and human metabolic diseases [69]. In a more specific example, a novel computational approach was applied to the GEM to identify biomarkers for inborn errors of metabolism [70]. This method revealed a set of 233 metabolites whose concentration is predicted to increase or decrease as a result of 176 possible dysfunctional enzymes. Human Recon 1 has also been used to explore tissue-specific metabolism across a number of major organ systems. The model was combined with tissue-specific gene expression data to predict tissue-specific activity of metabolic-disease genes and secreted metabolites [68]. An independently reconstructed human GEM [40] revealed the potential of systems modeling in human metabolism to aid in drug discovery [72]. Recently, efforts have focused on reconciling these reconstructions.

***Mus musculus***, the common laboratory mouse, has been found to have 99% similarity with the human genes in coding regions [73]. With extensive experimental data available, the mouse provides a terrific model organism for studying genetic systems of relevance to humans. The *M. musculus* metabolic GEM was used to simulate hybridoma cell line production of monoclonal antibodies (mAbs), and results were compared to cell culture data [74]. The model successfully predicted growth and production of lactate and ammonia, known byproducts of mammalian cell cultures that cause cell death and inhibit mAb synthesis. However, the model did not predict the production of a third commonly-observed byproduct, alanine, and did not explain the high production of lactate, ammonia, and alanine in animal cells. In 2009, an updated GEM was the subject of *in silico* analysis to identify strategies for optimizing cell density and mAb production in hybridoma cultures [75]. This GEM produced all expected amino acids. Based on cell culture measurements under various nutrient conditions and model simulations of internal metabolic states, potential feed-media conditions for enhancing cell density and mAb production were suggested.

# 4 GENOME-SCALE *IN SILICO* METABOLIC MODELS WITH APPLICATIONS IN BOTH INDUSTRIAL AND MEDICAL BIOTECHNOLOGY

## 4.1 *Mannheimia succiniciproducens* for succinate production

Succinate has importance in the food, agricultural, chemical, and pharmaceutical industries, and can be used in the synthesis of biodegradable polymers and green solvents. Currently, succinate is produced industrially from liquid petroleum gas via a chemical process.

To optimize microbial succinate production, a GEM was constructed for ***Mannheimia succiniciproducens*** [76]. Simulations of three mutant strains designed to increase succinate production were conducted. Good agreement was found between the experimental and *in silico* predictions for growth rate, and for succinate and acetate production by one strain. The prediction for formate was not in such good agreement, but the model accurately predicted that lactate, pyruvate and malate would not be produced. The flux simulation agreed with experiments in its prediction of the route to succinate. Two of the three mutants were more complex and results were initially not in agreement with experiment, but were similar after applying additional constraints to the model for one of the complex strains. The best succinate producing strains found using FBA yielded 92.59% of the maximum possible succinate production with only 25.50% reduction in growth rate. Constraint-based analysis on the model was later used to identify an ideal operating range of $CO_2$ to maximize cell growth rate and succinate production for a given glucose uptake rate [49].

### 4.2 *E. coli*

*E. coli* and *S. cerevisiae* (Section 4.3) are two of the best studied microbial species to date [77, 78], and serve as critically important organisms from which much about biology has been learned. The ***E. coli*** metabolic GEM has been extensively used in a wide spectrum of applications, including increased production of lycopene [79, 80], succinate [16, 81, 82], lactate [16, 83], malate [84], L-valine [85], L-threonine [86], additional amino acids [87], ethanol [88], hydrogen [17, 89], vanillin [17], and 1,3-propanediol (PDO) [16]. As one of the earliest GEMs and most extensively experimentally studied microorganisms, the *E. coli* model has been updated multiple times [90-92].

Quadruple gene deletions proposed by OptKnock were tested experimentally and resulted in a strain capable of increased lactate production (0.87-1.75 g/L per 2 g/L glucose) [83]. Adaptive evolution experiments on the engineered strains showed coupling of lactate production and growth, and the new strains increased lactate secretion rates. Constraint-based analysis of the *E. coli* GEM also guided strain design for increased production of the food additive malate, predicting the addition of a *M. succiniciproducens* reaction [84]. In another food engineering study, OptStrain identified three reactions to be introduced into *E. coli* for vanillin production [17]. OptKnock was then used to systematically search for gene deletions to enhance vanillin yield. For biofuels, an algorithm called OptReg [88] examined the effects of up-regulation of genes and to those of down-regulation and gene knock-outs to identify genes capable of enhancing ethanol production. With a focus on health applications, a MOMA-based procedure was used to sequentially examine and select sets of multiple gene deletions enabling optimal yields of the antioxidant lycopene, while still maintaining sufficient growth rates [79]. In another study, comparative genome analysis of *E. coli* and *M. succiniciproducens* was performed to predict five candidate genes to overproduce succinate in *E. coli* [81, 82].

A combination of strategies was used to develop an enhanced L-valine producing *E. coli* strain [85]. First, an L-valine producing strain was constructed by removing known feedback inhibition mechanisms and attenuation controls, and amplifying L-valine biosynthetic enzymes activity. This strain was improved in a stepwise manner using information derived from transcriptome profiling (i.e., the identification of a global regulator and exporter of L-valine). MOMA was applied to identify triple-knockout targets. The effect of the triple-knockout mutation was more drastic with respect to L-valine production in a strain overexpressing the global regulator gene and the exporter encoding gene than in a strain without these overexpressed genes. Analysis was also performed to uncover amplification targets for improved L-threonine production [86]. The strain was engineered to reduce byproduct accumulation during fed-batch culture by diverting the flux to L-threonine through overexpression of another GEM-identified gene.

Due to the inherent robustness of *E. coli* metabolism, only a subset of the metabolic genes was known to be lethal in single-gene deletion experiments [93]. Alternative approaches addressed this limitation by identifying synthetic lethals, or even higher-order lethal sets. These efforts dramatically expanded the range of knockout candidates. These lethal multiple-gene knockouts were identified *in silico* [42, 94-96], and antibacterial targets found through metabolite essentiality analyses [95] have been further explored [42].

### 4.3 *S. cerevisiae*

The yeast ***S. cerevisiae*** is one of the most widely studied model organisms for eukaryotes; research detailing its genetics, biochemistry, and physiology has provided a wealth of insight into mechanisms and behavior in higher-level organisms. *S. cerevisiae* is also capable of large-scale fermentation for the production of fuels, chemicals, pharmaceuticals, materials, nutritional compounds, and food ingredients.

The first *S. cerevisiae* GEM [97] was later expanded to establish a fully compartmentalized metabolic GEM [98], which was later updated [99]. An independently constructed GEM features a more detailed description of a lipid metabolism [100]. To reconcile the information in different models, a consensus GEM based on community knowledge has been collaboratively reconstructed, though this network reconstruction still lacks a biomass equation [101].

Ethanol is the predominant product in anaerobic fermentations with *S. cerevisiae*. With the availability of metabolic GEMs, constraint-based analyses can now be applied in new ways to systematically identify genetic engineering routes to increase ethanol production. For example, simulations predicted an insertion of the *gapn* gene as a strategy that could increase the ethanol yield, both with glucose as the sole carbon source as well as with a mixture of glucose and xylose, and experiments successfully validated this prediction [102]. Employing dFBA, another study demonstrated that the degree of compartmentalization in GEMs can impact the predicted mutant phenotypes [103].

Also in yeast, OptGene (an improvement of OptKnock) was used to identify potential metabolic engineering targets for increased production of succinate, glycerol, vanillin, and sesquiterpene [104, 105]. Growth phenotype predictions made using the *S. cerevisiae* GEM with simulated single-gene knockouts were consistent with experimental observations [106]. The phenotypic effects of multiple gene knockouts in the context of biological robustness and epistasis were also explored [107, 108].

As discussed in [109], such gene knockout studies can assist antimicrobial target discovery.

### 4.4 Aspergillus

Aspergillus is a filamentous fungus important to the medical and biotechnological (industrial and agricultural) communities. Aspergillus produces mycotoxins capable of contaminating crops, and can cause disease in immuno-compromised animals and humans. More constructively, Aspergillus are used in the production of bulk chemicals, enzymes, and pharmaceuticals. These applications have made Aspergillus a popular fungal species in research.

***Aspergillus niger*** is an industrial workhorse used to produce high yield products ranging from citrates and gluconates to important enzymes and proteins (e.g., human interferon). The *A. niger* metabolic GEM [110] was used to identify a gene deletion pair predicted to increase succinate production [111]. This mutant was tested experimentally, and as predicted, a significant increase in succinate production was observed when grown on both glucose and xylose. Unexpectedly, an increase in fumarate was seen when grown on xylose (though not when grown on glucose), suggesting that *A. niger* uses either the oxidative TCA cycle or the glyoxylate shunt for succinate production. *A. niger* converts up to 95% of the available carbon to organic acid, and, if unbuffered, can rapidly drop the pH to below 2. It has been found experimentally that depending on the ambient pH, *A. niger* produces a different organic acid. To study this process, the GEM was expanded to include information relating to the number of protons released for one mole of each acid. Using this GEM, the optimal strategy for acidifying the surrounding environment can be found computationally. The pH levels for citrate and oxalate were reproduced, verifying *in silico* the hypothesis that *A. niger* produces these to acidify its surrounding environment [112]. Other *Aspergillus* metabolic reconstructions include ***Aspergillus nidulans*** and ***Aspergillus oryzae*** (**Table 1)**. *A. nidulans* [113] is a model organism for studying cell development and gene regulation, and *A. oryzae* [114] has historically been used to produce soy sauce, miso and sake. *A. oryzae* is also used for the production of fungal enzymes such as alpha-amylase, glucoamylase, lipase and protease.

### 4.5 Host-Symbiote Relationships

The GEM-based study of host-symbiote relationships can shed light on the shared behavior and provide insight into industrial production abilities of the symbiote. Host-symbiote relationships have two primary modes of computational investigation: GEMs can be reconstructed for each participant and analyzed alone, or they can be constructed as an integrated network (see methanogens section). Obligate symbiotes in particular benefit from genome-scale *in silico* analysis as they cannot be cultured – and thus experimental results cannot be readily obtained.

***Buchnera aphidicola*** is an endosymbiote of the pea aphid whose metabolic GEM has been constructed to investigate symbiote-host interaction [115]. Interestingly, it was found that the *B. aphidicola* genome is essentially a subset of the *E. coli* genome [116]. Not surprisingly, a large percentage of genes from this network were predicted to be required for growth (84% by FBA and 95% by MOMA), showing that this organism's metabolic network is much less robust and complex than most. The bacterium cannot grow without secreting the essential amino acid, histidine, for use by its host. Further, the amount of essential amino acid produced by the bacterium *in silico* can be controlled by host supply of carbon and nitrogen substrates – possibly explaining the regulation of amino acid output to the host.

Another symbiotic bacterium, ***Rhizobium etli***, fixes atmospheric nitrogen into ammonium. A metabolic GEM for *R. etli* is of interest for plant development and in agriculture. *R. etli* obtains carbon sources from the plant and in turn provides ammonium, alanine and aspartate. Instead of using a biomass objective in its metabolic GEM [117], an objective function was formulated containing all compounds needed for symbiotic nitrogen fixation. This was done because the nitrogen fixation phase of the organism's life is of most interest, and in this phase does not grow. A double gene deletion was identified with a predicted increase in nitrogen fixation.

## 5 GENOME-SCALE *IN SILICO* REGULATORY MODELS

While genome-scale metabolic modeling strategies can be powerful, they are not completely predictive. In addition to incomplete or incorrect aspects of the reconstructions, one reason for failed predictions results from the lack of metabolite-level or transcriptional regulation of metabolism. Constraint-based analysis of metabolic GEMs typically assumes all metabolic enzymes are transcribed and available under all conditions, which is rarely the case. Thus, there is a compelling need to use procedures that incorporate metabolic regulation. Specifically, metabolic regulation can be categorized into two groups: transcriptional regulation that controls enzyme expression and metabolite-level regulation (e.g., allosteric regulation). **Figure 5** illustrates the interplay between both regulation types, such as when transcriptional regulation is itself affected by metabolite concentrations (e.g., feedback/feedforward inhibition/activation).

TRNs enhance metabolic simulations by providing information about transcriptionally active enzymes under different conditions. Recent efforts have attempted to reconstruct integrated networks, comprising both metabolic reactions and the regulatory rules that govern metabolic phenotypes, in order to more accurately represent metabolic phenotypes. One method linking the transcriptional state of an organism with metabolism is rFBA [3, 6]. rFBA uses Boolean rules to set gene activity for an enzyme as either ON or OFF based on the state of transcription factors and the environment.

The first integrated metabolic-regulatory network at the genome scale was reconstructed for *E. coli* [4]. This integrated model included 1,010 genes: 906 from the metabolic network [91], and 104 regulatory genes, whose products (i.e., transcription factors) together with other stimuli control the expression of 479 of the 906 metabolic enzymes and transports. The model predicted the outcomes of gene expression and growth phenotyping experiments, revealed knowledge gaps, and enabled the identification of additional components and interactions in each network. Steady-state regulatory FBA (SR-FBA), which reformulated rFBA using mixed integer linear programming [118], was later used to search the multiple solutions of rFBA rather than obtain only a particular flux state as with the Boolean-logic updating method for rFBA. More recently, a matrix formalism [119] was applied to the most updated integrated metabolic-regulatory model: among the 1260 genes in the metabolic model [92], 503 gene targets were regulated by the expression state of 125 transcription factors [120]. In addition to computing the transcription state of the genome, this formalism was used to describe intrinsic properties of the transcriptional regulatory states which could be analyzed by methods such as Monte Carlo sampling across a subset of all possible environments.

The first large-scale integrated metabolic-regulatory model in a eukaryotic organism was constructed for *S. cerevisiae* [6], containing 55 nutrient-regulated transcription factors that control a subset of the 750 genes in the metabolic network [98]. The rFBA approach [3] predicted gene expression changes and growth phenotypes of gene knockout strains.

## 6 CONCLUSIONS

We present an extensive review of the biotechnology applications of genome-scale modeling efforts to date, demonstrating the vast array of organisms already available for model-guided strain design and investigation of biochemical behavior. With the rise in high-throughput measurement technologies and the growing number of sequenced genomes, the continued construction of *in silico* GEMs will provide increasingly powerful tools to investigate biological systems. While existing models and corresponding analysis techniques have been developed primarily for metabolism, transcriptional regulation and transcription-translation processes are emerging. Many of the studies highlighted herein reveal the utility of GEMs for generating predictions for experimental testing and design, as well as providing valuable insight into metabolic function. Commonly, *in silico* simulations are employed to study the effects of genetic perturbations on the stoichiometric abilities of a cell. In this way, these studies have used GEMs to predict engineering strategies to enhance properties of interest in an organism and/or inhibit harmful mechanisms of pathogens or in disease.

Looking forward, technology and computational method development will continuecontinues to improve the predictive capability and usefulness of *in silico* GEMs. These efforts focus on "integration," whether in regard to heterogeneous high-throughput data types, or different scales and scopes of biological processes. As technological advances enable increasingly comprehensive and accurate measurements of intracellular and extracellular metabolite concentrations [121], these data will greatly inform GEM reconstruction and analysis, including for dynamics. Integration of cellular regulation and signaling with metabolic information is important for predicting diverse network states. The successes with the genome-scale TRN in *E. coli* [4, 5, 120] and *S. cerevisiae* [6] demonstrate the potential of metabolic networks controlled by gene expression. Similarly, the recent invention of a genome-scale transcriptional-translational network model (demonstrated in *E. coli* [7]) will allow for integrated analysis of transcriptomic and metabolic states. Contrary to these successes though, analysis of allosteric regulation between metabolites and enzymes is still lacking in GEMS because of sparse high-throughput data and applicable computational methods to uncover these genome-scale relationships. Similar technical problems exist for intracellular signaling networks as GEMs, although integration of metabolic, transcriptional regulatory, and signaling networks has been investigated [5, 122]. As network integration becomes commonplace, consistent formatting and naming conventions must become a priority to assist in seamless melding of information. Achieving integration of the different biochemical processes will open another avenue to ultimately realize whole-cell simulation.

## ACKNOWLEDGEMENTS

The authors would like to acknowledge financial support for this work from an NSF CAREER Award (NDP) and a Howard Temin Pathway to Independence Award in Cancer Research (NDP) from NCI. P.-J.K. was supported by the IGB Postdoctoral Fellows Program.

## 7 REFERENCES

[1] Schilling, C. H., Palsson, B. O., Assessment of the metabolic capabilities of Haemophilus influenzae Rd through a genome-scale pathway analysis. *J. Theor. Biol.* 2000, *203*, 249-283.

[2] Feist, A. M., Herrgard, M. J., Thiele, I., Reed, J. L., Palsson, B. O., Reconstruction of biochemical networks in microorganisms. *Nat. Rev. Microbiol.* 2009, *7*, 129-143.

[3] Covert, M. W., Schilling, C. H., Palsson, B., Regulation of gene expression in flux balance models of metabolism. *J. Theor. Biol.* 2001, *213*, 73-88.

[4] Covert, M. W., Knight, E. M., Reed, J. L., Herrgard, M. J., Palsson, B. O., Integrating high-throughput and computational data elucidates bacterial networks. *Nature* 2004, *429*, 92-96.

[5] Covert, M. W., Xiao, N., Chen, T. J., Karr, J. R., Integrating metabolic, transcriptional regulatory and signal transduction models in Escherichia coli. *Bioinformatics* 2008, *24,* 2044-2050.

[6] Herrgard, M. J., Lee, B. S., Portnoy, V., Palsson, B. O., Integrated analysis of regulatory and metabolic networks reveals novel regulatory mechanisms in Saccharomyces cerevisiae. *Genome Res.* 2006, *16*, 627-635.

[7] Thiele, I., Jamshidi, N., Fleming, R. M., Palsson, B. O., Genome-scale reconstruction of Escherichia coli's transcriptional and translational machinery: a

knowledge base, its mathematical formulation, and its functional characterization. *PLoS Comput. Biol.* 2009, *5*, e1000312.

[8] Park, J. M., Kim, T. Y., Lee, S. Y., Constraints-based genome-scale metabolic simulation for systems metabolic engineering. *Biotechnol. Adv.* 2009.

[9] Durot, M., Bourguignon, P. Y., Schachter, V., Genome-scale models of bacterial metabolism: reconstruction and applications. *FEMS Microbiol. Rev.* 2009, *33*, 164-190.

[10] Kim, H. U., Kim, T. Y., Lee, S. Y., Metabolic flux analysis and metabolic engineering of microorganisms. *Mol. BioSyst.* 2008, *4*, 113-120.

[11] Price, N. D., Reed, J. L., Palsson, B. O., Genome-scale models of microbial cells: evaluating the consequences of constraints. *Nat. Rev. Microbiol.* 2004, *2*, 886-897.

[12] Kauffman, K. J., Prakash, P., Edwards, J. S., Advances in flux balance analysis. *Curr. Opin. Biotechnol.* 2003, *14*, 491-496.

[13] Edwards, J. S., Ibarra, R. U., Palsson, B. O., In silico predictions of Escherichia coli metabolic capabilities are consistent with experimental data. *Nat. Biotechnol.* 2001, *19*, 125-130.

[14] Mahadevan, R., Edwards, J. S., Doyle, F. J., 3rd, Dynamic flux balance analysis of diauxic growth in Escherichia coli. *Biophys. J.* 2002, *83*, 1331-1340.

[15] Segre, D., Vitkup, D., Church, G. M., Analysis of optimality in natural and perturbed metabolic networks. *Proc. Natl. Acad. Sci. U. S. A.* 2002, *99*, 15112-15117.

[16] Burgard, A. P., Pharkya, P., Maranas, C. D., Optknock: a bilevel programming framework for identifying gene knockout strategies for microbial strain optimization. *Biotechnol. Bioeng.* 2003, *84*, 647-657.

[17] Pharkya, P., Burgard, A. P., Maranas, C. D., OptStrain: a computational framework for redesign of microbial production systems. *Genome Res.* 2004, *14*, 2367-2376.

[18] Zhu, Y., Zhang, Y., Li, Y., Understanding the industrial application potential of lactic acid bacteria through genomics. *Appl. Microbiol. Biotechnol.* 2009, *83*, 597-610.

[19] Teusink, B., Wiersma, A., Jacobs, L., Notebaart, R. A., Smid, E. J., Understanding the adaptive growth strategy of Lactobacillus plantarum by in silico optimisation. *PLoS Comput. Biol.* 2009, *5*, e1000410.

[20] Oliveira, A. P., Nielsen, J., Forster, J., Modeling Lactococcus lactis using a genome-scale flux model. *BMC Microbiol.* 2005, *5*, 39.

[21] Oddone, G. M., Mills, D. A., Block, D. E., A dynamic, genome-scale flux model of Lactococcus lactis to increase specific recombinant protein expression. *Metab. Eng.* 2009.

[22] Pastink, M. I., Teusink, B., Hols, P., Visser, S., *et al.*, Genome-scale model of Streptococcus thermophilus LMG18311 for metabolic comparison of lactic acid bacteria. *Appl. Environ. Microbiol.* 2009, *75*, 3627-3633.

[23] Nogales, J., Palsson, B. O., Thiele, I., A genome-scale metabolic reconstruction of Pseudomonas putida KT2440: iJN746 as a cell factory. *BMC Syst. Biol.* 2008, *2*, 79.

[24] Puchalka, J., Oberhardt, M. A., Godinho, M., Bielecka, A., *et al.*, Genome-scale reconstruction and analysis of the Pseudomonas putida KT2440 metabolic network facilitates applications in biotechnology. *PLoS Comput. Biol.* 2008, *4*, e1000210.

[25] Lee, S. K., Chou, H., Ham, T. S., Lee, T. S., Keasling, J. D., Metabolic engineering of microorganisms for biofuels production: from bugs to synthetic biology to fuels. *Curr. Opin. Biotechnol.* 2008, *19*, 556-563.

[26] Senger, R. S., Papoutsakis, E. T., Genome-scale model for Clostridium acetobutylicum: Part I. Metabolic network resolution and analysis. *Biotechnol. Bioeng.* 2008, *101*, 1036-1052.

[27] Lee, J., Yun, H., Feist, A. M., Palsson, B. O., Lee, S. Y., Genome-scale reconstruction and in silico analysis of the Clostridium acetobutylicum ATCC 824 metabolic network. *Appl. Microbiol. Biotechnol.* 2008, *80*, 849-862.

[28] Feist, A. M., Scholten, J. C., Palsson, B. O., Brockman, F. J., Ideker, T., Modeling methanogenesis with a genome-scale metabolic reconstruction of Methanosarcina barkeri. *Mol. Syst. Biol.* 2006, *2*, 2006 0004.

[29] Stolyar, S., Van Dien, S., Hillesland, K. L., Pinel, N., *et al.*, Metabolic modeling of a mutualistic microbial community. *Mol. Syst. Biol.* 2007, *3*, 92.

[30] de Lorenzo, V., Systems biology approaches to bioremediation. *Curr. Opin. Biotechnol.* 2008, *19*, 579-589.

[31] Durot, M., Le Fevre, F., de Berardinis, V., Kreimeyer, A., *et al.*, Iterative reconstruction of a global metabolic model of Acinetobacter baylyi ADP1 using high-throughput growth phenotype and gene essentiality data. *BMC Syst. Biol.* 2008, *2*, 85.

[32] Sun, J., Sayyar, B., Butler, J. E., Pharkya, P., *et al.*, Genome-scale constraint-based modeling of Geobacter metallireducens. *BMC Syst. Biol.* 2009, *3*, 15.

[33] Mahadevan, R., Bond, D. R., Butler, J. E., Esteve-Nunez, A., *et al.*, Characterization of metabolism in the Fe(III)-reducing organism Geobacter sulfurreducens by constraint-based modeling. *Appl. Environ. Microbiol.* 2006, *72*, 1558-1568.

[34] Izallalen, M., Mahadevan, R., Burgard, A., Postier, B., *et al.*, Geobacter sulfurreducens strain engineered for increased rates of respiration. *Metab. Eng.* 2008, *10*, 267-275.

[35] Radrich, K., Tsuruoka, Y., Dobson, P., Gevorgyan, A., *et al.*, Reconstruction of an in silico metabolic model of Arabidopsis thaliana through database integration. *Available from Nature Precedings* 2009.

[36] Manichaikul, A., Ghamsari, L., Hom, E. F., Lin, C., *et al.*, Metabolic network analysis integrated with transcript verification for sequenced genomes. *Nat. Methods* 2009, *6*, 589-592.

[37] Gonzalez, O., Gronau, S., Falb, M., Pfeiffer, F., *et al.*, Reconstruction, modeling & analysis of Halobacterium salinarum R-1 metabolism. *Mol. BioSyst.* 2008, *4*, 148-159.

[38] Fu, P., Genome-scale modeling of Synechocystis sp. PCC 6803 and prediction of pathway insertion. *Journal of Chemical Technology & Biotechnology* 2009, *84*, 473-483.

[39] Duarte, N. C., Becker, S. A., Jamshidi, N., Thiele, I., *et al.*, Global reconstruction of the human metabolic network based on genomic and bibliomic data. *Proc. Natl. Acad. Sci. U. S. A.* 2007, *104*, 1777-1782.

[40] Ma, H., Sorokin, A., Mazein, A., Selkov, A., *et al.*, The Edinburgh human metabolic network reconstruction and its functional analysis. *Mol. Syst. Biol.* 2007, *3*, 135.

[41] Becker, S. A., Palsson, B. O., Genome-scale reconstruction of the metabolic network in Staphylococcus aureus N315: an initial draft to the two-dimensional annotation. *BMC Microbiol.* 2005, *5*, 8.

[42] Kim, T. Y., Kim, H. U., Lee, S. Y., Metabolite-centric approaches for the discovery of antibacterials using genome-scale metabolic networks. *Metab. Eng.* 2009.

[43] Heinemann, M., Kummel, A., Ruinatscha, R., Panke, S., In silico genome-scale reconstruction and validation of the Staphylococcus aureus metabolic network. *Biotechnol. Bioeng.* 2005, *92*, 850-864.

[44] Lee, D. S., Burd, H., Liu, J., Almaas, E., *et al.*, Comparative genome-scale metabolic reconstruction and flux balance analysis of multiple Staphylococcus aureus genomes identify novel antimicrobial drug targets. *J. Bacteriol.* 2009, *191*, 4015-4024.

[45] Raghunathan, A., Price, N. D., Galperin, M. Y., Makarova, K. S., *et al.*, In Silico Metabolic Model and Protein Expression of Haemophilus influenzae Strain Rd KW20 in Rich Medium. *OMICS* 2004, *8*, 25-41.

[46] Beste, D. J., Hooper, T., Stewart, G., Bonde, B., *et al.*, GSMN-TB: a web-based genome-scale network model of Mycobacterium tuberculosis metabolism. *Genome Biol.* 2007, *8*, R89.

[47] Jamshidi, N., Palsson, B. O., Investigating the metabolic capabilities of Mycobacterium tuberculosis H37Rv using the in silico strain iNJ661 and proposing alternative drug targets. *BMC Syst. Biol.* 2007, *1*, 26.

[48] Colijn, C., Brandes, A., Zucker, J., Lun, D. S., *et al.*, Interpreting expression data with metabolic flux models: predicting Mycobacterium tuberculosis mycolic acid production. *PLoS Comput. Biol.* 2009, *5*, e1000489.

[49] Kim, T. Y., Kim, H. U., Song, H., Lee, S. Y., In silico analysis of the effects of H2 and CO2 on the metabolism of a capnophilic bacterium Mannheimia succiniciproducens. *J. Biotechnol.* 2009, *In Press, Corrected Proof.*

[50] Raman, K., Yeturu, K., Chandra, N., targetTB: a target identification pipeline for Mycobacterium tuberculosis through an interactome, reactome and genome-scale structural analysis. *BMC Syst. Biol.* 2008, *2*, 109.

[51] Oberhardt, M. A., Puchalka, J., Fryer, K. E., Martins dos Santos, V. A., Papin, J. A., Genome-scale metabolic network analysis of the opportunistic pathogen Pseudomonas aeruginosa PAO1. *J. Bacteriol.* 2008, *190*, 2790-2803.

[52] Schilling, C. H., Covert, M. W., Famili, I., Church, G. M., *et al.*, Genome-scale metabolic model of Helicobacter pylori 26695. *J. Bacteriol.* 2002, *184*, 4582-4593.

[53] Thiele, I., Vo, T. D., Price, N. D., Palsson, B. O., Expanded metabolic reconstruction of Helicobacter pylori (iIT341 GSM/GPR): an in silico genome-scale characterization of single- and double-deletion mutants. *J. Bacteriol.* 2005, *187*, 5818-5830.

[54] Raghunathan, A., Reed, J., Shin, S., Palsson, B., Daefler, S., Constraint-based analysis of metabolic capacity of Salmonella typhimurium during host-pathogen interaction. *BMC Syst. Biol.* 2009, *3*, 38.

[55] Abuoun, M., Suthers, P. F., Jones, G. I., Carter, B. R., *et al.*, Genome scale reconstruction of a Salmonella metabolic model: comparison of similarity and differences with a commensal Escherichia coli strain. *J. Biol. Chem.* 2009.

[56] Baart, G. J., Zomer, B., de Haan, A., van der Pol, L. A., *et al.*, Modeling Neisseria meningitidis metabolism: from genome to metabolic fluxes. *Genome Biol.* 2007, *8*, R136.

[57] Navid, A., Almaas, E., Genome-scale reconstruction of the metabolic network in Yersinia pestis, strain 91001. *Mol. BioSyst.* 2009, *5*, 368-375.

[58] Chavali, A. K., Whittemore, J. D., Eddy, J. A., Williams, K. T., Papin, J. A., Systems analysis of metabolism in the pathogenic trypanosomatid Leishmania major. *Mol. Syst. Biol.* 2008, *4*, 177.

[59] Suthers, P. F., Dasika, M. S., Kumar, V. S., Denisov, G., *et al.*, A genome-scale metabolic reconstruction of Mycoplasma genitalium, iPS189. *PLoS Comput. Biol.* 2009, *5,* e1000285.

[60] Mazumdar, V., Snitkin, E. S., Amar, S., Segre, D., Metabolic network model of a human oral pathogen. *J. Bacteriol.* 2009, *191*, 74-90.

[61] Kjeldsen, K. R., Nielsen, J., In silico genome-scale reconstruction and validation of the Corynebacterium glutamicum metabolic network. *Biotechnol. Bioeng.* 2009, *102,* 583-597.

[62] Shinfuku, Y., Sorpitiporn, N., Sono, M., Furusawa, C., *et al.*, Development and experimental verification of a genome-scale metabolic model for Corynebacterium glutamicum. *Microb. Cell Fact.* 2009, *8,* 43.

[63] Oh, Y. K., Palsson, B. O., Park, S. M., Schilling, C. H., Mahadevan, R., Genome-scale reconstruction of metabolic network in Bacillus subtilis based on high-throughput phenotyping and gene essentiality data. *J. Biol. Chem.* 2007, *282*, 28791-28799.

[64] Henry, C. S., Zinner, J. F., Cohoon, M. P., Stevens, R. L., iBsu1103: a new genome-scale metabolic model of Bacillus subtilis based on SEED annotations. *Genome Biol.* 2009, *10*, R69.

[65] Borodina, I., Krabben, P., Nielsen, J., Genome-scale analysis of Streptomyces coelicolor A3(2) metabolism. *Genome Res.* 2005, *15*, 820-829.

[66] Borodina, I., Siebring, J., Zhang, J., Smith, C. P., *et al.*, Antibiotic overproduction in Streptomyces coelicolor A3 2 mediated by phosphofructokinase deletion. *J. Biol. Chem.* 2008, *283*, 25186-25199.

[67] Khannapho, C., Zhao, H., Bonde, B. K., Kierzek, A. M., *et al.*, Selection of objective function in genome scale flux balance analysis for process feed development in antibiotic production. *Metab. Eng.* 2008, *10*, 227-233.

[68] Shlomi, T., Cabili, M. N., Herrgard, M. J., Palsson, B. O., Ruppin, E., Network-based prediction of human tissue-specific metabolism. *Nat. Biotechnol.* 2008, *26*, 1003-1010.

[69] Lee, D. S., Park, J., Kay, K. A., Christakis, N. A., *et al.*, The implications of human metabolic network topology for disease comorbidity. *Proc. Natl. Acad. Sci. U. S. A.* 2008, *105*, 9880-9885.

[70] Shlomi, T., Cabili, M. N., Ruppin, E., Predicting metabolic biomarkers of human inborn errors of metabolism. *Mol. Syst. Biol.* 2009, *5*, 263.

[71] Butler, M., Animal cell cultures: recent achievements and perspectives in the production of biopharmaceuticals. *Appl. Microbiol. Biotechnol.* 2005, *68*, 283-291.

[72] Ma, H., Goryanin, I., Human metabolic network reconstruction and its impact on drug discovery and development. *Drug Discov. Today* 2008, *13*, 402-408.

[73] Waterston, R. H., Lindblad-Toh, K., Birney, E., Rogers, J., *et al.*, Initial sequencing and comparative analysis of the mouse genome. *Nature* 2002, *420*, 520-562.

[74] Sheikh, K., Forster, J., Nielsen, L. K., Modeling hybridoma cell metabolism using a generic genome-scale metabolic model of Mus musculus. *Biotechnol. Prog.* 2005, *21*, 112-121.

[75] Selvarasu, S., Wong, V. V., Karimi, I. A., Lee, D. Y., Elucidation of metabolism in hybridoma cells grown in fed-batch culture by genome-scale modeling. *Biotechnol. Bioeng.* 2009, *102*, 1494-1504.

[76] Kim, T. Y., Kim, H. U., Park, J. M., Song, H., *et al.*, Genome-scale analysis of Mannheimia succiniciproducens metabolism. *Biotechnol. Bioeng.* 2007, *97*, 657-671.

[77] Feist, A. M., Palsson, B. O., The growing scope of applications of genome-scale metabolic reconstructions using Escherichia coli. *Nat. Biotechnol.* 2008, *26*, 659-667.

[78] Nielsen, J., Jewett, M. C., Impact of systems biology on metabolic engineering of Saccharomyces cerevisiae. *FEMS Yeast Res.* 2008, *8*, 122-131.

[79] Alper, H., Jin, Y. S., Moxley, J. F., Stephanopoulos, G., Identifying gene targets for the metabolic engineering of lycopene biosynthesis in Escherichia coli. *Metab. Eng.* 2005, *7*, 155-164.

[80] Alper, H., Miyaoku, K., Stephanopoulos, G., Construction of lycopene-overproducing E. coli strains by combining systematic and combinatorial gene knockout targets. *Nat. Biotechnol.* 2005, *23*, 612-616.

[81] Lee, S. J., Lee, D. Y., Kim, T. Y., Kim, B. H., *et al.*, Metabolic engineering of Escherichia coli for enhanced production of succinic acid, based on genome comparison and in silico gene knockout simulation. *Appl. Environ. Microbiol.* 2005, *71*, 7880-7887.

[82] Wang, Q., Chen, X., Yang, Y., Zhao, X., Genome-scale in silico aided metabolic analysis and flux comparisons of Escherichia coli to improve succinate production. *Appl. Microbiol. Biotechnol.* 2006, *73*, 887-894.

[83] Fong, S. S., Burgard, A. P., Herring, C. D., Knight, E. M., *et al.*, In silico design and adaptive evolution of Escherichia coli for production of lactic acid. *Biotechnol. Bioeng.* 2005, *91*, 643-648.

[84] Moon, S. Y., Hong, S. H., Kim, T. Y., Lee, S. Y., Metabolic engineering of Escherichia coli for the production of malic acid. *Biochem. Eng. J.* 2008, *40*, 312-320.

[85] Park, J. H., Lee, K. H., Kim, T. Y., Lee, S. Y., Metabolic engineering of Escherichia coli for the production of L-valine based on transcriptome analysis and in silico gene knockout simulation. *Proc. Natl. Acad. Sci. U. S. A.* 2007, *104*, 7797-7802.

[86] Lee, K. H., Park, J. H., Kim, T. Y., Kim, H. U., Lee, S. Y., Systems metabolic engineering of Escherichia coli for L-threonine production. *Mol. Syst. Biol.* 2007, *3*, 149.

[87] Pharkya, P., Burgard, A. P., Maranas, C. D., Exploring the overproduction of amino acids using the bilevel optimization framework OptKnock. *Biotechnol. Bioeng.* 2003, *84*, 887-899.

[88] Pharkya, P., Maranas, C. D., An optimization framework for identifying reaction activation/inhibition or elimination candidates for overproduction in microbial systems. *Metab. Eng.* 2006, *8*, 1-13.

[89] Jones, P. R., Improving fermentative biomass-derived H2-production by engineering microbial metabolism. *Int. J. Hydrogen Energy* 2008, *33*, 5122-5130.

[90] Edwards, J. S., Palsson, B. O., The Escherichia coli MG1655 in silico metabolic genotype: its definition, characteristics, and capabilities. *Proc. Natl. Acad. Sci. U. S. A.* 2000, *97*, 5528-5533.

[91] Reed, J. L., Vo, T. D., Schilling, C. H., Palsson, B. O., An expanded genome-scale model of Escherichia coli K-12 (iJR904 GSM/GPR). *Genome Biol.* 2003, *4*, R54.

[92] Feist, A. M., Henry, C. S., Reed, J. L., Krummenacker, M., *et al.*, A genome-scale metabolic reconstruction for Escherichia coli K-12 MG1655 that accounts for 1260 ORFs and thermodynamic information. *Mol. Syst. Biol.* 2007, *3*, 121.

[93] Gerdes, S. Y., Scholle, M. D., Campbell, J. W., Balazsi, G., *et al.*, Experimental determination and system level analysis of essential genes in Escherichia coli MG1655. *J. Bacteriol.* 2003, *185*, 5673-5684.

[94] Suthers, P. F., Zomorrodi, A., Maranas, C. D., Genome-scale gene/reaction essentiality and synthetic lethality analysis. *Mol. Syst. Biol.* 2009, *5*, 301.

[95] Kim, P. J., Lee, D. Y., Kim, T. Y., Lee, K. H., *et al.*, Metabolite essentiality elucidates robustness of Escherichia coli metabolism. *Proc. Natl. Acad. Sci. U. S. A.* 2007, *104*, 13638-13642.

[96] Ghim, C. M., Goh, K. I., Kahng, B., Lethality and synthetic lethality in the genome-wide metabolic network of Escherichia coli. *J. Theor. Biol.* 2005, *237*, 401-411.

[97] Forster, J., Famili, I., Fu, P., Palsson, B. O., Nielsen, J., Genome-scale reconstruction of the Saccharomyces cerevisiae metabolic network. *Genome Res.* 2003, *13*, 244-253.

[98] Duarte, N. C., Herrgard, M. J., Palsson, B. O., Reconstruction and validation of Saccharomyces cerevisiae iND750, a fully compartmentalized genome-scale metabolic model. *Genome Res.* 2004, *14*, 1298-1309.

[99] Mo, M. L., Palsson, B. O., Herrgard, M. J., Connecting extracellular metabolomic measurements to intracellular flux states in yeast. *BMC Syst. Biol.* 2009, *3*, 37.

[100] Nookaew, I., Jewett, M. C., Meechai, A., Thammarongtham, C., *et al.*, The genome-scale metabolic model iIN800 of Saccharomyces cerevisiae and its validation: a scaffold to query lipid metabolism. *BMC Syst. Biol.* 2008, *2*, 71.

[101] Herrgard, M. J., Swainston, N., Dobson, P., Dunn, W. B., *et al.*, A consensus yeast metabolic network reconstruction obtained from a community approach to systems biology. *Nat. Biotechnol.* 2008, *26*, 1155-1160.

[102] Bro, C., Regenberg, B., Forster, J., Nielsen, J., In silico aided metabolic engineering of Saccharomyces cerevisiae for improved bioethanol production. *Metab. Eng.* 2006, *8*, 102-111.

[103] Hjersted, J. L., Henson, M. A., Steady-state and dynamic flux balance analysis of ethanol production by Saccharomyces cerevisiae. *IET Syst. Biol.* 2009, *3*, 167-179.

[104] Patil, K. R., Rocha, I., Forster, J., Nielsen, J., Evolutionary programming as a platform for in silico metabolic engineering. *BMC Bioinformatics* 2005, *6*, 308.

[105] Asadollahi, M. A., Maury, J., Patil, K. R., Schalk, M., *et al.*, Enhancing sesquiterpene production in Saccharomyces cerevisiae through in silico driven metabolic engineering. *Metab. Eng.* 2009.

[106] Famili, I., Forster, J., Nielsen, J., Palsson, B. O., Saccharomyces cerevisiae phenotypes can be predicted by using constraint-based analysis of a genome-scale reconstructed metabolic network. *Proc. Natl. Acad. Sci. U. S. A.* 2003, *100*, 13134-13139.

[107] Deutscher, D., Meilijson, I., Kupiec, M., Ruppin, E., Multiple knockout analysis of genetic robustness in the yeast metabolic network. *Nat. Genet.* 2006, *38*, 993-998.

[108] Segre, D., Deluna, A., Church, G. M., Kishony, R., Modular epistasis in yeast metabolism. *Nat. Genet.* 2005, *37*, 77-83.

[109] Trawick, J. D., Schilling, C. H., Use of constraint-based modeling for the prediction and validation of antimicrobial targets. *Biochem. Pharmacol.* 2006, *71*, 1026-1035.

[110] Andersen, M. R., Nielsen, M. L., Nielsen, J., Metabolic model integration of the bibliome, genome, metabolome and reactome of Aspergillus niger. *Mol. Syst. Biol.* 2008, *4*, 178.

[111] Meijer, S., Nielsen, M. L., Olsson, L., Nielsen, J., Gene deletion of cytosolic ATP: citrate lyase leads to altered organic acid production in Aspergillus niger. *J. Ind. Microbiol. Biotechnol.* 2009.

[112] Andersen, M. R., Lehmann, L., Nielsen, J., Systemic analysis of the response of Aspergillus niger to ambient pH. *Genome Biol.* 2009, *10*, R47.

[113] David, H., Ozcelik, I. S., Hofmann, G., Nielsen, J., Analysis of Aspergillus nidulans metabolism at the genome-scale. *BMC Genomics* 2008, *9*, 163.

[114] Vongsangnak, W., Olsen, P., Hansen, K., Krogsgaard, S., Nielsen, J., Improved annotation through genome-scale metabolic modeling of Aspergillus oryzae. *BMC Genomics* 2008, *9*, 245.

[115] Thomas, G. H., Zucker, J., Macdonald, S. J., Sorokin, A., *et al.*, A fragile metabolic network adapted for cooperation in the symbiotic bacterium Buchnera aphidicola. *BMC Syst. Biol.* 2009, *3*, 24.

[116] Pal, C., Papp, B., Lercher, M. J., Csermely, P., *et al.*, Chance and necessity in the evolution of minimal metabolic networks. *Nature* 2006, *440*, 667-670.

[117] Resendis-Antonio, O., Reed, J. L., Encarnacion, S., Collado-Vides, J., Palsson, B. O., Metabolic reconstruction and modeling of nitrogen fixation in Rhizobium etli. *PLoS Comput. Biol.* 2007, *3*, 1887-1895.

[118] Shlomi, T., Eisenberg, Y., Sharan, R., Ruppin, E., A genome-scale computational study of the interplay between transcriptional regulation and metabolism. *Mol. Syst. Biol.* 2007, *3*, 101.

[119] Gianchandani, E. P., Papin, J. A., Price, N. D., Joyce, A. R., Palsson, B. O., Matrix formalism to describe functional states of transcriptional regulatory systems. *PLoS Comput. Biol.* 2006, *2*, e101.

[120] Gianchandani, E. P., Joyce, A. R., Palsson, B. O., Papin, J. A., Functional States of the genome-scale Escherichia coli transcriptional regulatory system. *PLoS Comput. Biol.* 2009, *5*, e1000403.

[121] Kell, D. B., Brown, M., Davey, H. M., Dunn, W. B., *et al.*, Metabolic footprinting and systems biology: the medium is the message. *Nat. Rev. Microbiol.* 2005, *3*, 557-565.

[122] Lee, J. M., Gianchandani, E. P., Eddy, J. A., Papin, J. A., Dynamic analysis of integrated signaling, metabolic, and regulatory networks. *PLoS Comput. Biol.* 2008, *4*, e1000086.

[123] Teusink, B., Wiersma, A., Molenaar, D., Francke, C., *et al.*, Analysis of growth of Lactobacillus plantarum WCFS1 on a complex medium using a genome-scale metabolic model. *J. Biol. Chem.* 2006, *281*, 40041-40048.

[124] Vo, T. D., Paul Lee, W. N., Palsson, B. O., Systems analysis of energy metabolism elucidates the affected respiratory chain complex in Leigh's syndrome. *Mol. Genet. Metab.* 2007, *91*, 15-22.

## TABLES

**Table 1.** Genome-scale metabolic models to date. Under "Domain", bacteria, eukaryote, and archaea are marked as "b", "e", and "a", respectively.

| Organism | Domain | Model Details<br># rxns / # mets / # genes | Refs | Demonstrated/Intended Applications |
|---|---|---|---|---|
| *Lactobacillus plantarum* | b | 643 / 531 / 721 | [19] | lactate [123] |
| *Lactococcus lactis* | b | 621 / 422 / 358 | [20] | lactate [20], diacetyl [20] |
| *Streptococcus thermophilus* | b | 522 / --- / 429 | [22] | lactate, acetaldehyde |
| *Pseudomonas putida* | b | 950 / 911 / 746 | [23] | polyhydroxyalkanoates [23, 24], bioremediation, biocatalytic chemicals, improvement of fossil fuel quality, promoting plant growth, pest control |
| *Pseudomonas putida* | | 877 / 886 / 815 | [24] | |
| *Clostridium acetobutylicum* | b | 502 / 479 / 432 | [27] | acetone, butanol, ethanol, hydrogen |
| *Clostridium acetobutylicum* | | 552 / 422 / 474 | [26] | |
| *Methanosarcina barkeri* | a | 509 / 558 / 692 | [28] | methane |
| *Desulfovibrio vulgaris* | b | --- / --- / --- | [29] | methane |
| *Methanococcus maripaludis* | a | --- / --- / --- | [29] | methane |
| *Acinetobacter baylyi* | b | 875 / 701 / 774 | [31] | pollutant degradation, lipases, proteases, bioemulsifiers, cyanophycine, various biopolymers |
| *Geobacter metallireducens* | b | 697 / 769 / 747 | [32] | reducing Fe(III), bioremediation of uranium, plutonium, technetium & vadium, |
| | | | | fuel cell development |
| *Geobacter sulfurreducens* | b | 523 / 541 / 588 | [33] | reducing Fe(III), bioremediation of uranium, plutonium, technetium & vadium [34], fuel cell development [34] |
| *Arabidopsis thaliana* | e | --- / --- / --- | [35] | photosynthetic plant cell, various secondary metabolites, flavonoid, polyamine metabolism |
| *Chlamydomonas reinhardtii* | e | 259 / 113 / 174 | [36] | photosynthetic green algae, hydrogen production |
| *Halobacterium salinarum* | a | 711 / 557 / 490 | [37] | producing bacteriorhodopsin |
| *Synechocystis sp* | b | 831 / 704 / 633 | [38] | photosynthetic cyanobateria, ethanol production [38] |
| *Staphylococcus aureus* | b | 640 / 571 / 691 | [41] | antibiotic target [41-44] |
| *Staphylococcus aureus* | | 774 / 712 / 551 | [43] | |
| *Staphylococcus aureus* (multiple strains) | | 1444~97 / 1399~1437 / 522~47 | [44] | |
| *Haemophilus influenzae* | b | 461 / 451 / 412 | [1] | antibiotic target [1, 45] |
| *Pseudomonas aeruginosa* | b | 883 / 760 / 1056 | [51] | antibiotic target [51] |
| *Mycobacterium tuberculosis* | b | 849 / 739 / 726 | [46] | antibiotic target [42, 46-48, 50] |
| *Mycobacterium tuberculosis* | | 939 / 828 / 661 | [47] | |
| *Helicobacter pylori* | b | 476 / 485 / 341 | [53] | antibiotic target [42, 53] |
| *Salmonella typhimurium* | b | 1087 / 744 / 1038 | [54] | antibiotic target [54] |
| *Salmonella typhimurium* | | 1964 / 1036 / 945 | | |
| *Neisseria meningitidis* | b | 496 / 471 / 555 | [56] | vaccine development, antibiotic target |
| *Yersinia pestis* | b | 1020 / 825 / 818 | [57] | vaccine development, antibiotic target |
| *Leishmania major* | e | 1112 / 1101 / 560 | [58] | antibiotic target [58] |
| *Mycoplasma genitalium* | b | 262 / 274 / 189 | [59] | antibiotic target [59] |
| *Porphyromonas gingivalis* | b | 679 / 564 / --- | [60] | control of negative inflammatory responses [60] |
| *Corynebacterium glutamicum*<br>*Corynebacterium glutamicum* | b | 446 / 411 / 446<br>502 / 423 / 277 | [61]<br>[62] | lactic and succinate [62], L-lysine [61], glutamate, ethanol |
| *Bacillus subtilis* | b | 1437 / 1138 / 1103 | [64] | antibiotics, industrial enzymes and proteins, nucleosides and vitamins |
| *Streptomyces coelicolor* | b | 971 / 500 / 711 | [65] | secondary metabolites (antibiotics, immunosuppressants, anti-cancer agents) [66, 67] |
| *Homo sapiens*<br>*Homo sapiens* | e | 3311 / 2766 / 1496<br>2823 / 2671 / 2322 | [39]<br>[40] | biomarker of inborn error [70], understanding disease comorbidity toward diagnosis and prevention [69], identification of mutations causing defects in Leigh's cells [124], predicting tissue-specific activity of metabolic genes [68] |
| *Mus musculus* | e | 1344 / 1042 / --- | [75] | mouse hybridoma cells for enhanced production of monoclonal antibodies [75] |
| *Mannheimia succiniciproducens* | b | 686 / 519 / 425 | [76] | succinate [76] |
| *Escherichia coli* | b | 2077 / 1039 / 1260 | [92] | lycopene [79, 80], succinate [16, 81, 82], lactate [16, 83], malate [84], L-valine [85], L-threonine [86], additional amino acids [87], ethanol [88], hydrogen [17, 89], vanillin [17], 1,3-propanediol (PDO) [16], gene KO [94-96], antibacterial target [42] |
| *Saccharomyces cerevisiae* (fully-compartmentalized) | e | 1412 / 1228 / 904 | [99] | ethanol [102, 103], succinate [104], glycerol [104], vanillin [104], sesquiterpene [105], gene KO [106-108] |
| *Saccharomyces cerevisiae* (lipids emphasized) | | 1446 / 1013 / 800 | [100] | |
| *Saccharomyces cerevisiae* (consensus model) | | 1857 / 1168 / 832 | [101] | |
| *Aspergillus niger* | e | 1197 / 1045 / 871 | [112] | succinate [111], citrates and oxalates [112], additional organic acids, industrial enzymes, proteins (chymosin, human interferon) |
| *Aspergillus nidulans* | e | 676 / 733 / 666 | [113] | model organism for studies of development biology & gene regulation, sharing many applicative properties of *A. niger* |
| *Aspergillus oryzae* | e | 1053 / 1073 / 1314 | [114] | fermented sauces, industrial enzymes |
| *Buchnera aphidicola* | b | 263 / 240 / 196 | [115] | symbiotes producing histidine [115] |
| *Rhizobium etli* | b | 387 / 371 / 363 | [117] | symbiotic nitrogen fixation [117] |

## FIGURE LEGENDS

**Figure 1.** Completed genome sequences and genome-scale models (GEMs) available to date.

**Figure 2.** Applications of GEMs in industrial and medical biotechnology.

**Figure 3.** Iterative process of model generation, hypothesis formation, and model refinement to guide strain design for enhanced microbial production.

**Figure 4.** Iterative modeling of pathogens to identify new antibiotic targets and therapeutic strategies.

**Figure 5.** Overview of key processes governing the interplay between metabolic and transcriptional regulatory networks, demonstrating the utility of integrated modeling.

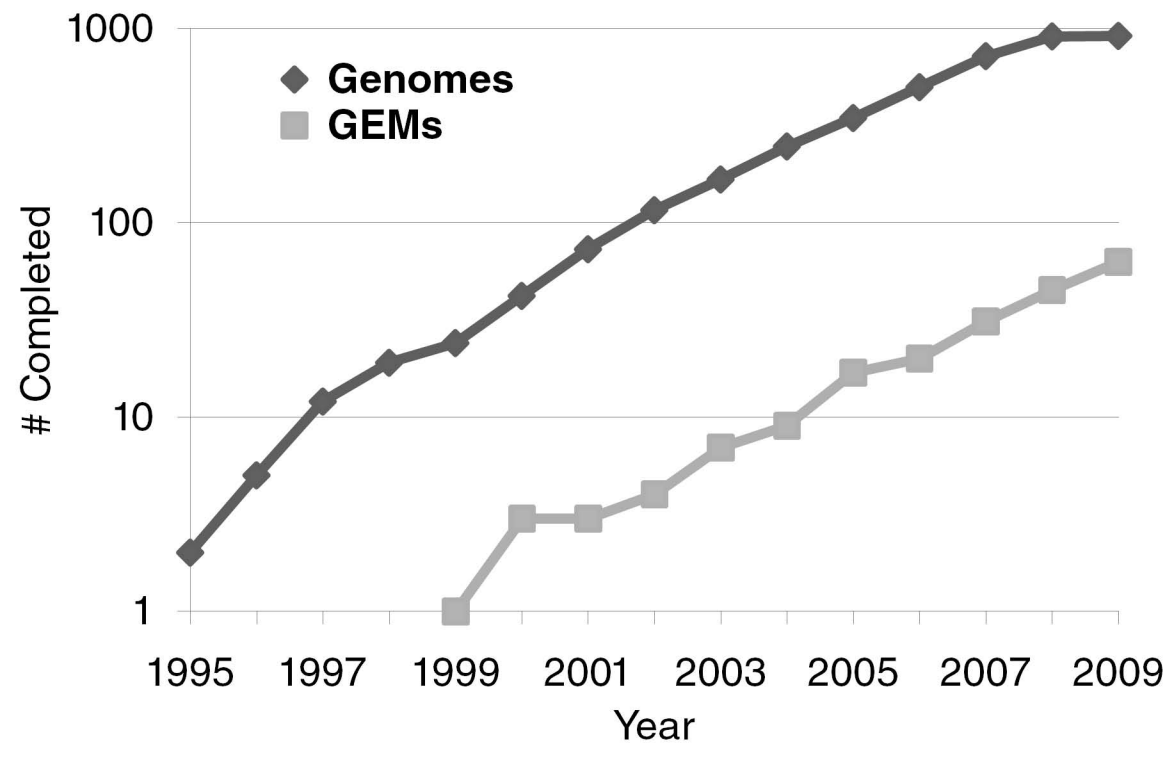

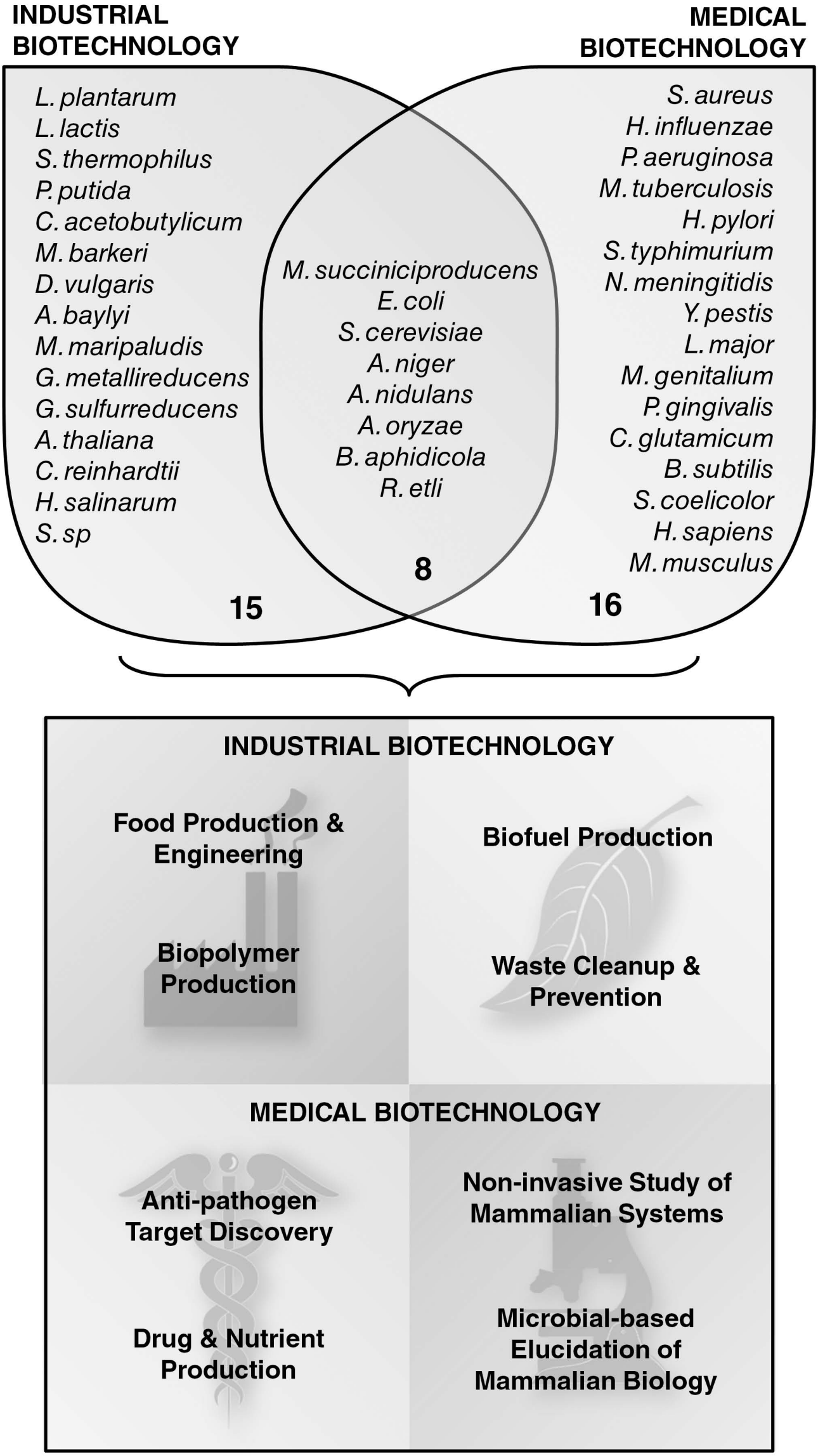

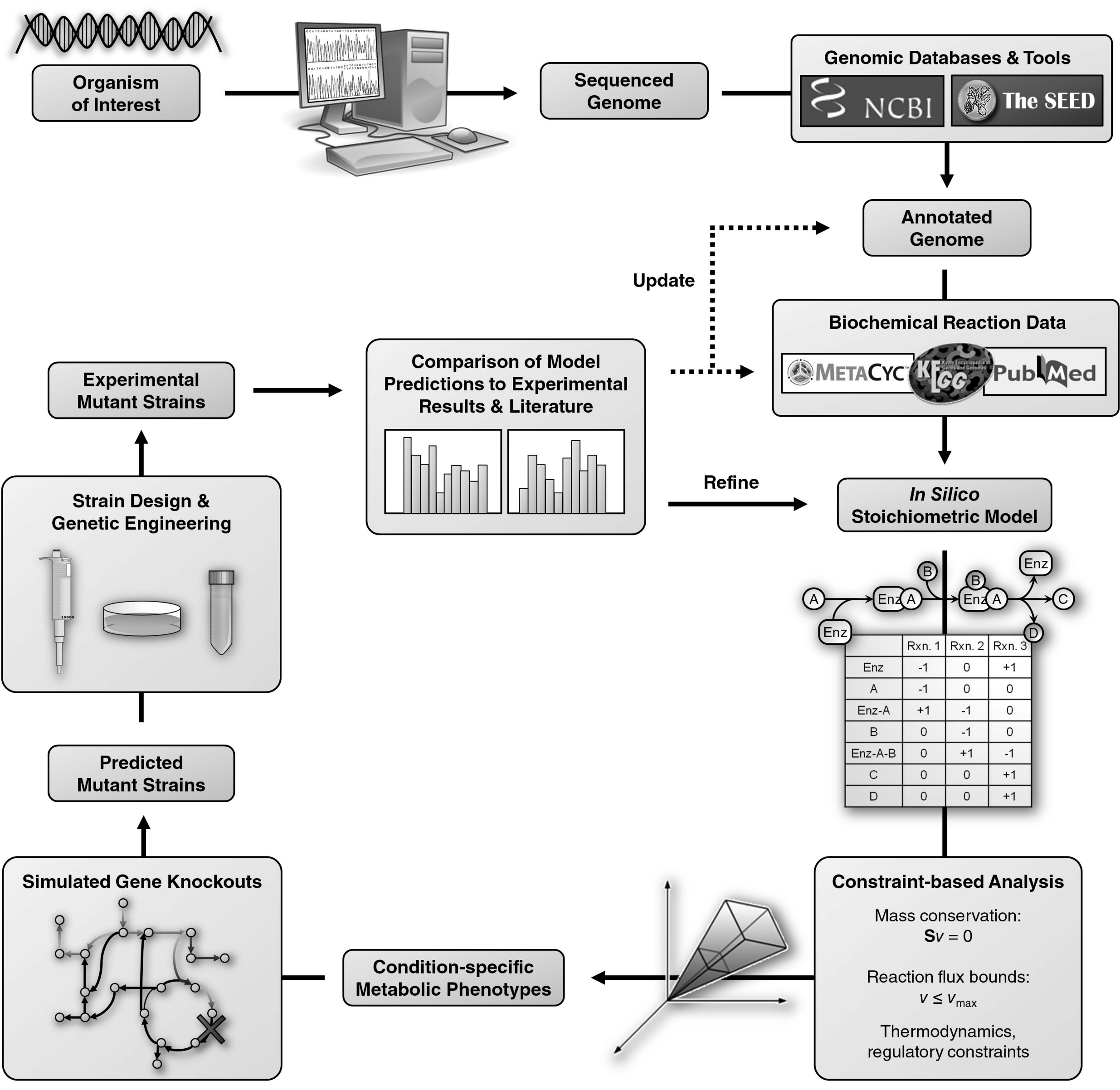
Organism of Interest
Sequenced Genome
Genomic Databases & Tools
NCBI
The SEED
Annotated Genome
Update
Biochemical Reaction Data
METACYC
KEGG
PubMed
Experimental Mutant Strains
Comparison of Model Predictions to Experimental Results & Literature
Refine
In Silico Stoichiometric Model
Strain Design & Genetic Engineering
Predicted Mutant Strains
A
Enz
Enz A
B
Enz A
C
D
| | Rxn. 1 | Rxn. 2 | Rxn. 3 |
| Enz | -1 | 0 | +1 |
| A | -1 | 0 | 0 |
| Enz-A | +1 | -1 | 0 |
| B | 0 | -1 | 0 |
| Enz-A-B | 0 | +1 | -1 |
| C | 0 | 0 | +1 |
| D | 0 | 0 | +1 |
Simulated Gene Knockouts
Condition-specific Metabolic Phenotypes
Constraint-based Analysis
Mass conservation:
$Sv = 0$
Reaction flux bounds:
$v \leq v_{max}$
Thermodynamics, regulatory constraints

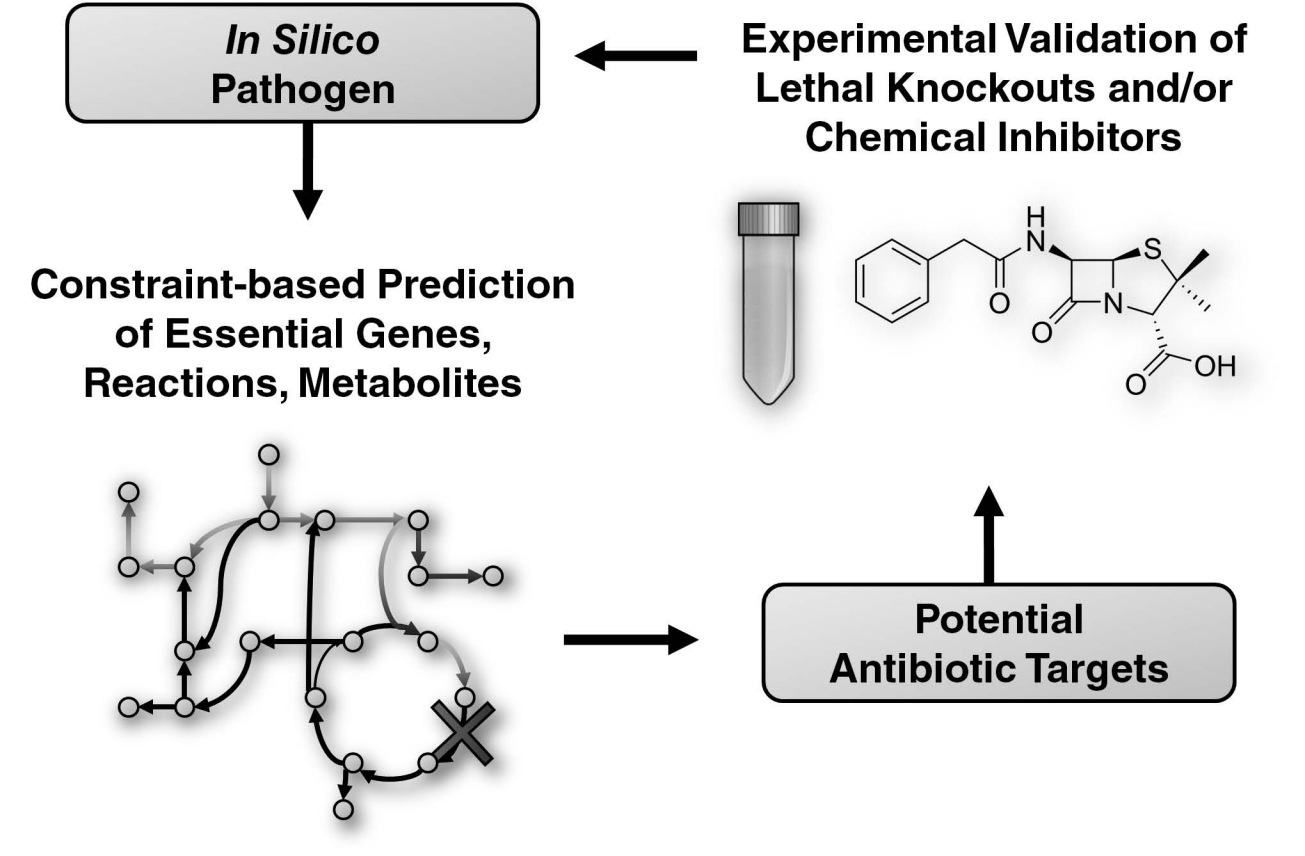

In Silico Pathogen
Experimental Validation of Lethal Knockouts and/or Chemical Inhibitors
Constraint-based Prediction of Essential Genes, Reactions, Metabolites
Potential Antibiotic Targets
H
N
S
N
O
O
O
O
OH

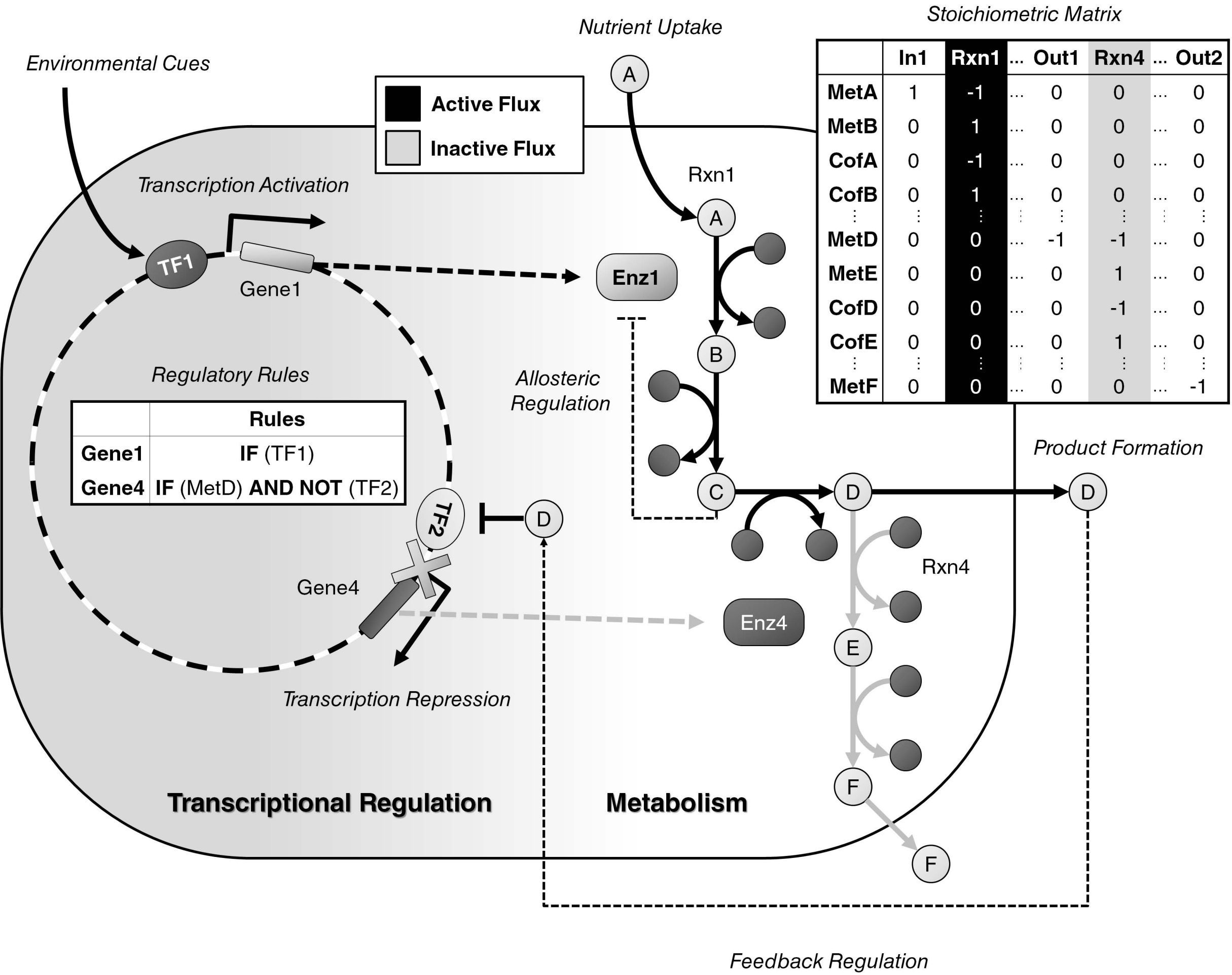

Environmental Cues
Nutrient Uptake
Stoichiometric Matrix
| | In1 | Rxn1 | ... | Out1 | Rxn4 | ... | Out2 |
| MetA | 1 | -1 | ... | 0 | 0 | ... | 0 |
| MetB | 0 | 1 | ... | 0 | 0 | ... | 0 |
| CofA | 0 | -1 | ... | 0 | 0 | ... | 0 |
| CofB | 0 | 1 | ... | 0 | 0 | ... | 0 |
| MetD | 0 | 0 | ... | -1 | -1 | ... | 0 |
| MetE | 0 | 0 | ... | 0 | 1 | ... | 0 |
| CofD | 0 | 0 | ... | 0 | -1 | ... | 0 |
| CofE | 0 | 0 | ... | 0 | 1 | ... | 0 |
| MetF | 0 | 0 | ... | 0 | 0 | ... | -1 |
Active Flux
Inactive Flux
Transcription Activation
TF1
Gene1
Enz1
Rxn1
A
B
C
D
E
F
Allosteric Regulation
Regulatory Rules
| | Rules |
| Gene1 | IF (TF1) |
| Gene4 | IF (MetD) AND NOT (TF2) |
TF2
Gene4
Enz4
Rxn4
Product Formation
Transcription Repression
Transcriptional Regulation
Metabolism
Feedback Regulation